\documentclass[aps, pra, amsmath, floatfix, reprint]{revtex4-1}
\usepackage{blindtext}
\usepackage{physics}
\usepackage{graphicx}
\usepackage{hyperref}

\usepackage[caption=false]{subfig} 
\begin{document}
\newcommand{\mvcomment}[1]{{\textbf{#1}}}

\title{Scalable Hardware-Efficient Qubit Control with Single Flux Quantum Pulse
    Sequences}
\author{Kangbo Li}
\affiliation{Department of Physics, University of Wisconsin-Madison, Madison,
    WI 53706, USA}
\author{R. McDermott}
\affiliation{Department of Physics, University of Wisconsin-Madison, Madison,
    WI 53706, USA}
\author{Maxim G. Vavilov}
\affiliation{Department of Physics, University of Wisconsin-Madison, Madison,
    WI 53706, USA}
\date{\today}

\begin{abstract}
    The hardware overhead associated with microwave
    control is a major obstacle to scale-up of superconducting quantum
    computing.
    An alternative approach involves irradiation of the qubits with trains
    of Single Flux Quantum (SFQ) pulses,
    pulses of voltage whose time integral is precisely equal to the
    superconducting flux quantum.
    Here we describe the  derivation and validation of compact SFQ pulse
    sequences in which
    classical bits are clocked  to the qubit at a frequency that is roughly
    a factor 5 higher than the qubit oscillation
    frequency, allowing for variable
    pulse-to-pulse timing. The control sequences are constructed by
    repeated
    streaming of short subsequence registers that are designed to suppress
    leakage out of the computational manifold. With a single global clock,
    high-fidelity ($>99.99$\%) control of qubits resonating at over 20
    distinct frequencies is possible.
    SFQ pulses can be stored locally and delivered to the qubits via a
    proximal classical Josephson digital circuit,
    offering the possibility of a streamlined,
    low-footprint classical coprocessor for
    monitoring errors and feeding back to the qubit array.
\end{abstract}

\maketitle

\section{INTRODUCTION \label{sec:INTRODUCTION}}

\par A fault-tolerant quantum computer will possess a
computational power far exceeding that
of any classical computer \cite{Feynman1982},
and superconducting integrated circuits are a promising
physical platform for the realization of scalable qubits \cite{Clarke2008}.
While the error thresholds of the two-dimensional surface code are within reach
\cite{Fowler2012, Barends2014}, quantum error detection involves a massive
hardware overhead: estimates suggest that a general-purpose fault-tolerant
quantum computer will require millions of physical qubits, far
beyond current capabilities.
Conventionally, qubit control
pulses are generated by single-sideband modulation of a
microwave carrier tone; accurate control of both the in-phase and quadrature
pulse amplitudes allows arbitrary rotations on
the Bloch sphere \cite{Motzoi2009, Lucero2010, Chow2010, Chen2016}.
In order to minimize crosstalk between neighboring qubit channels, it is
generally necessary to arrange the qubit array so that devices are biased at a
handful of distinct operating frequencies; this approach makes it possible to
address a large-scale multiqubit circuit with a small number of carrier tones,
resulting in a significant hardware savings \cite{Asaad2016}. In addition,
there have been proposals to recycle pulse waveforms across the qubit array
\cite{Versluis2017}. However, the control waveform that is delivered to the
qubit is the convolution of the applied waveform with the transfer function of
the wiring in the qubit cryostat, which in general is not well controlled. As
wiring transfer functions can vary substantially across the array, it is not
clear that recycling of control waveforms will allow high-fidelity control.
Moreover, the separate high-bandwidth control lines for each qubit channel
entail a massive heat load on the millikelvin stage. Finally, the significant
latency associated with the round trip from the quantum array to the
room-temperature classical coprocessor will limit the performance of any scheme
to use high-fidelity projective measurement and feedback to stabilize the
qubits \cite{vijay2012, Campagne-Ibarcq2013, Riste2012}.

\par An alternative approach is to control the quantum array using a classical
coprocessor that is integrated tightly with the qubits at the millikelvin
stage.
Recently we proposed an approach to coherent control involving irradiation of
the qubit with trains of quantized flux pulses derived from the Single Flux
Quantum (SFQ) digital logic family. Here, classical bits of information are
stored as the presence or absence of a phase slip across a Josephson junction
in a given clock cycle \cite{Likharev1991}; the phase slip results in a voltage
pulse whose time integral is precisely quantized to $\Phi_0 = h/2e$, the
superconducting flux quantum. For typical parameters, SFQ pulse amplitudes are
of order 1~mV and pulse durations are around 2~ps, roughly two orders of
magnitude shorter than the typical qubit oscillation period. As a result, the
SFQ pulse imparts a delta
function-like kick to the qubit that induces a coherent rotation in the qubit
subspace \cite{McDermott2014}. In the first experimental implementation of this
idea, gate fidelity was limited by spurious quasiparticle generation by the
dissipative SFQ pulse driver, which was integrated on the same
chip as the qubit circuit \cite{Leonard2019}. In next-generation devices, it is
expected that segregation of classical control elements and quantum elements on
the two separate chips of a multichip module (MCM) will lead to a significant
suppression of quasiparticle poisoning \cite{McDermott2018}. Ultimately, the
fidelity of naive, resonant SFQ pulse trains will be limited by leakage out
of the computational subspace, with achievable fidelity around $99.9\%$ for
typical values	of qubit anharmonicity and gate times around 20~ns. This
fidelity is likely insufficient for fault-tolerant operations in a large-scale
surface code array.

\par In order to achieve SFQ-based gates with fidelity well beyond threshold,
it is possible to clock SFQ bits to the qubit at a higher rate, allowing  for
SFQ pulse trains with variable pulse-to-pulse separation. In previous work,
genetic algorithms were used to derive optimized pulse sequences that lead to
very low leakage and gate fidelities better than 99.99\% \cite{Liebermann2016}.
However, the number of bits required to achieve high fidelity was rather high,
and the genetic approach provides no intuition as to why a particular sequence
yields good performance. While this proof-of-principle demonstration suggests
that SFQ control sequences might offer a low-footprint alternative to
conventional microwave sequences, the following questions remain:

\begin{itemize}
    \item What is the minimum number of classical bits
          needed to control a qubit to a given level of fidelity?
    \item Is it possible to achieve high-fidelity control of qubits
          at different frequencies using a single global clock, as
          would be ideal for the surface code?
\end{itemize}

\par In this work, we describe a method to derive hardware-efficient SFQ
control sequences for scalable qubit control: SCALable Leakage Optimized Pulse
Sequences (SCALLOPS). The sequences are built up from short subsequences
consisting of
35-55 classical bits that are repeatedly streamed to the qubit. Leakage is
minimized at the subsequence level; because the subsequences are short,
it is possible to perform efficient search over the subsequence space in order
to optimize gate fidelity. For SFQ clock frequency roughly a factor 5 greater
than the characteristic qubit frequency, we achieve high-fidelity qubit
rotations for a large number of discrete qubit frequencies, as required for
low-crosstalk control of a large-scale qubit array designed to implement the
surface code.

\par This manuscript is organized as follows. In Section \ref{sec:CQED} we
present a simple analytic framework for the study of SFQ-based pulse sequences.
In Section \ref{sec:SCALLOPS} we introduce the key features of the SCALLOPS
approach: (1) a palindromic sequence construction consisting of symmetric pulse
pairs that suppresses errors within the qubit subspace; (2) accurate control of
qubits resonating at a number of discrete frequencies by repeated streaming of
a short pulse subsequence; and (3) a normal graph algorithm to minimize leakage
from the computational subspace at the subsequence level.
Finally, in Section \ref{sec:CONCLUSION} we conclude and provide
a perspective on the hardware requirements for the implementation of SCALLOPS.

\par Throughout this paper, we consider a fixed SFQ clock frequency of 25~GHz,
so that SFQ pulses are delivered to the qubit at intervals that are integer
multiples of the 40~ps clock period. In addition, we consider transmon qubits
with fixed anharmonicity $(\omega_{10}-\omega_{21})/2\pi$ of 250~MHz, where
$\omega_{10} \equiv \omega_{q}$ is the qubit transition frequency and
$\omega_{21}$ is the transition frequency between the qubit $\ket{1}$ state and
the noncomputational $\ket{2}$ state. Finally, for the sake of concreteness we
target high fidelity for a single gate,
the $Y_{\pi/2}$ rotation; the SCALLOPS approach is readily generalized to
arbitrary single-qubit sequences.

\section{Model for SFQ Control of a Transmon Qubit \label{sec:CQED}}

\subsection{General Model \label{sec:CQED:GENERAL}}

\begin{figure}[t!]
    \includegraphics[scale=0.8]{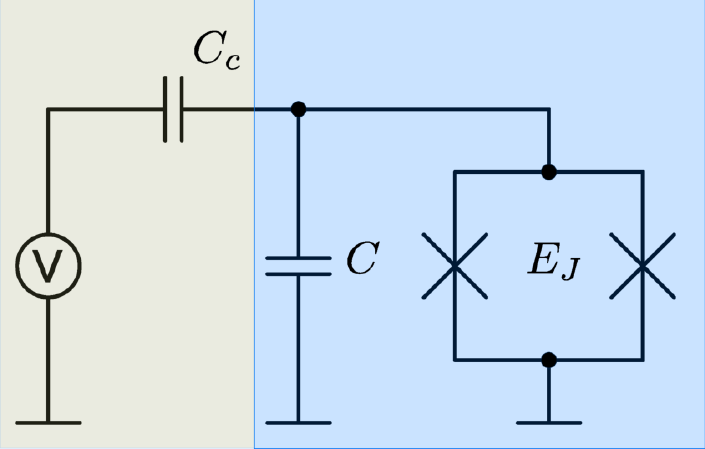}
    \caption{(Color online) SFQ drive circuit (yellow) capacitively coupled
        to a
        transmon qubit (blue).
        \label{fig:sfq_transmon}}
\end{figure}

\par We consider a conventional
transmon qubit \cite{Koch2007} coupled via a small capacitance $C_c$
to an SFQ driver modeled as a time-dependent voltage source $V_{\rm SFQ}(t)$,
as drawn in Fig. \ref{fig:sfq_transmon}.
The Hamiltonian of the undriven transmon is written as
\begin{equation}
    H_{\rm fr} = \frac{\hat{Q}^2}{2C'} -E_J \cos\hat{\varphi},
    \label{eqn:hfree}
\end{equation}
where $\hat Q$ and $\hat\varphi$ are the charge and phase operators of the
transmon,
$E_J$ is the transmon Josephson energy, and $C' = C_c + C$ is the sum of the
coupling
capacitance $C_c$ and the transmon self capacitance $C$.
$H_{\rm fr}$ can be diagonalized in closed form,
with the resulting energy eigenfunctions
$\bra{\varphi}\ket{\gamma}$ and energies $E_{\gamma}$
represented by the Mathieu functions and coefficients.

\par Interaction between the transmon and the SFQ pulse driver
adds the following term to the Hamiltonian:
\begin{equation}
    H_{\rm SFQ}(t) = \frac{C_c}{C'} V_{\rm SFQ}(t) \hat{Q},
    \label{eqn:hsfq}
\end{equation}
where the SFQ pulse $V_{\rm SFQ}(t)$ satisfies the condition
$\int_{-\infty}^{\infty} V_{\rm SFQ}(t)\mathrm{d} t = \Phi_0$.
Since the pulse width (around 2~ps) is much less than the Larmor period
(around 200~ps) of the transmon,
we model the SFQ pulse as a Dirac delta function: $V_{\rm SFQ}(t) =
    \Phi_0\delta(t)$.
The charge operator can be constructed in the basis of $\ket{\gamma}$ once
$H_{\rm fr}$ is diagonalized.
The free evolution of the transmon then becomes
\begin{equation}
    U_{\rm fr}(t) = \exp\left( -\frac{i}{\hbar} \sum
    \ket{\gamma}\bra{\gamma}
    E_{\gamma}t \right).
\end{equation}
The time evolution operator for a transmon subjected to an SFQ pulse is
\cite{McDermott2014}
\begin{equation}
    U_{\rm SFQ} = \exp\left(-i \Phi_0 (C_c/C') \hat Q\right).
\end{equation}

\subsection{Three-level Model \label{sec:CQED:THREELEVEL}}

\par It is advantageous to restrict the size of the transmon Hilbert space in
order to accelerate the search for high-fidelity pulse sequences. Since leakage
outside the computational subspace is dominated by population of the first
noncomputational state $\ket{2}$, we truncate the transmon to a three-level
qutrit; subsequent sequence validation will be performed on a more complete
model of the transmon consisting of 7 states. Within the three-level subspace,
the operators $H_{\rm fr}$ and $H_{\rm SFQ}(t)$ take the form
\begin{align}
     & H^{(3)}_{\rm fr} = \frac{\hbar \omega_{q}}{2} \hat{\Sigma}_z, \quad
    H^{(3)}_{\rm SFQ}(t) = \frac{\hbar \omega_y(t)}{2} \hat{\Sigma}_y;
    \label{eqn:hmatrices}
    \\
     & \hat{\Sigma}_z =
    \begin{bmatrix}
        0 & 0 & 0       \\
        0 & 2 & 0       \\
        0 & 0 & 4-2\eta
    \end{bmatrix}, \quad
    \hat{\Sigma}_y = i
    \begin{bmatrix}
        0 & -1      & 0        \\
        1 & 0       & -\lambda \\
        0 & \lambda & 0
    \end{bmatrix}.
\end{align}
Here, we have introduced the notation $\omega_y(t) = -\left(2 V(t)/\hbar\right)
    (C_c/C') \bra{1}\hat{Q}\ket{0}$;
$\eta= 1-\omega_{21}/\omega_{q}$ represents the fractional anharmonicity of the
transmon; and
$\lambda = \bra{2}\hat{Q}\ket{1}/\bra{1}\hat{Q}\ket{0}$.

\par Next, we derive the three-level matrix form for the free evolution
of the transmon and for the evolution of the transmon subjected to a single SFQ
pulse. The free evolution is given by the diagonal matrix $U^{(3)}_{\rm fr}(t)
    = \exp \left(-i \omega_q t \hat{\Sigma}_z / 2\right)$.
In the qubit subspace, the effect of $U^{(3)}_{\rm fr}$ is clearly a
precession at the rate $\omega_{q}$. For the time evolution under a single SFQ
pulse, we can write
\begin{align}
    U^{(3)}_{\rm SFQ} & = \exp\left(-\frac{i\hat{\Sigma}_y}{2}
    \int_{-\infty}^{\infty}
    \omega_y(t) dt \right) =
    \exp\left(\frac{-i\delta \theta \hat{\Sigma}_y}{2}\right)
    \label{eqn:sfq_pulse_three_level},
\end{align}
where
\begin{equation}
    \delta \theta = \int_{-\infty}^{\infty}\omega_y(t)dt =(2 \Phi_0/\hbar)
    (C_c/C')
    \bra{1}\hat{Q}\ket{0}
    \label{eqn:tip_angle}
\end{equation}
is the tip angle associated with a single SFQ pulse.
Using the Cayley-Hamilton theorem on $\hat{\Sigma}_y$,
we obtain the property $\hat{\Sigma}_y^{3} = (\lambda^2+1)\hat{\Sigma}_y$,
which yields
$\hat{\Sigma}_y^n = (\lambda^2+1)^{\frac{n-2}{2}}\hat{\Sigma}_y^2 $
for even $n$ and
$\hat{\Sigma}_y^n = (\lambda^2+1)^{\frac{n-1}{2}}\hat{\Sigma}_y$
for odd $n$.
We then expand and regroup Eq. \eqref{eqn:sfq_pulse_three_level}:
\begin{equation}
    \begin{split}
        U^{(3)}_{\rm SFQ} =  \hat 1 \,\,+ &\sum_{\text{even}, n\geq
            2}^{\infty}\frac{(\lambda^2+1)^
            {\frac{n-2}{2}}(-i \delta \theta/2)^n}{n!}
        \hat{\Sigma}_y^2 \\
        + &\sum_{\text{odd}, n\geq
            1}^{\infty}\frac{(\lambda^2+1)^{\frac{n-1}{2}}
            (-i \delta \theta/2)^n}{n!} \hat{\Sigma}_y.
    \end{split}
    \label{eqn:sfq_expansion}
\end{equation}
The two sums in Eq.~\eqref{eqn:sfq_expansion} yield
\begin{equation}
    \begin{split}
        &U^{(3)}_{\rm SFQ} =  \frac{1}{\kappa^2}  \times  \\
        &\begin{bmatrix}
            \lambda ^2+\cos \left(\kappa \delta \theta  /2 \right)
                      &
            -\kappa \sin \left( \kappa \delta  \theta  /2\right)
                      &
            2 \lambda  \sin ^2\left(\kappa \delta \theta /4\right)
            \\
            \kappa \sin \left( \kappa \delta  \theta  /2\right)
                      &
            \kappa^2 \cos \left(\kappa \delta \theta  /2 \right)
                      &
            -\kappa \lambda  \sin \left(\kappa \delta \theta
            \right/2)   \\
            2 \lambda  \sin ^2\left(\kappa \delta \theta /4\right)
                      &
            \kappa \lambda	\sin \left(\kappa \delta \theta
            \right/2) &
            1+ \lambda ^2\cos \left(\kappa \delta \theta  /2
            \right)     \\
        \end{bmatrix}, \label{eqn:u_sfq_matrix}
    \end{split}
\end{equation}
where $\kappa = \sqrt{\lambda ^2+1}$.

\par To see the effect of a single SFQ pulse, we compare this time evolution to
a $y$-rotation by angle $\delta\theta$ in the qubit subspace:
\begin{equation}
    Y_{\delta \theta} =
    \begin{bmatrix}
        \cos\left( \delta\theta/2 \right) &
        -\sin\left( \delta\theta/2 \right)  \\
        \sin\left( \delta\theta/2 \right) &
        \cos\left( \delta\theta/2 \right)
    \end{bmatrix} \label{eqn:ideal_matrix}.
\end{equation}
We observe that: (1) within the three-level model,
the SFQ pulse provides a rotation in the qubit subspace
that is slightly smaller than $\delta \theta$; and
(2) leakage from state $\ket{1}$ to state $\ket{2}$ is first order in $\delta
    \theta$, while leakage from state $\ket{0}$ to state $\ket{2}$ is
second order
in $\delta \theta$.

\subsection{Pulse Sequences and Gate Fidelity \label{sec:CQED:SEQUENCE}}

The above analysis shows that it is impossible to perform coherent qubit
rotations with a single SFQ pulse without incurring significant excitation of
noncomputational states. However, composite sequences consisting of multiple
SFQ pulses spaced by appropriate time intervals can achieve low leakage and
high gate fidelity. More specifically, we consider a high-speed SFQ clock that
delivers pulses to
the transmon according to a vector of binary variables $\mathbf{S}$,
where $S_i = 0$ if no SFQ pulse is applied
on the $i^{th}$ clock edge and $S_i = 1$ if an SFQ pulse is applied.
Using these expressions, the total time evolution operator of the gate $U_{\rm
            G}$,
time ordered in terms of clock edges, can be written as
\begin{equation}
    U_{\rm G} = \mathcal{T} \left\{ \prod_{i}^{N_c}\left( \delta_{S_i 1}
    U_{\rm
            fr}(T_c) U_{\rm SFQ}
    + \delta_{S_i 0} U_{\rm fr}(T_c)\right) \right\}.
    \label{eqn:evolution}
\end{equation}
Here, $N_c$ is the number of clock cycles in the sequence and $T_c$ is the
clock period.

\par We evaluate the fidelity of the gate $U_{\rm G}$ as in \cite{Bowdrey2002};

\begin{equation}
    \begin{split}
        F_{\rm avg} &= \frac{1}{6}\sum_{\ket{\alpha}\in {\cal V}}
        \left| \bra{\alpha} U_G^{\dagger} Y_{\pi/2}\ket{\alpha}
        \right|^2,
        \label{eqn:fid}
    \end{split}
\end{equation}
where the summation runs over the six states ${\cal V}$ aligned along the
cardinal directions of the Bloch sphere
\begin{align}
    \ket{x_\pm} & =
    \frac{\ket{0}\pm\ket{1}}{\sqrt{2}} ,               \\ \nonumber
    \ket{y_\pm} & =
    \frac{\ket{0}\pm i\ket{1}}{\sqrt{2}} ,             \\ \nonumber
    \ket{z_+}   & = \ket{0}, \,\, \ket{z_-} = \ket{1};
\end{align}
and the gate $Y_{\pi/2}$ gate is represented by the following matrix in
the qubit subspace:
\begin{equation}
    Y_{\pi/2} = \frac{1}{\sqrt{2}}\left(\dyad{0}{0} + \dyad{1}{1} +
    \dyad{1}{0}-\dyad{0}{1}\right).
\end{equation}
The crux of the problem then becomes proper selection of $\mathbf{S}$
so that $U_{\rm G}$ becomes a high-fidelity $Y_{\pi/2}$ gate.

\section{SCALLOP sequences \label{sec:SCALLOPS}}

\subsection{Symmetric SFQ Pulse Pairs \label{sec:SCALLOPS:SYMMETRIC}}
\begin{figure}[t!]
    \centering
    \includegraphics{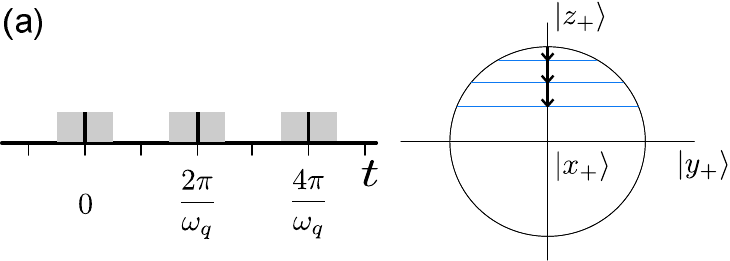}\\
    \includegraphics{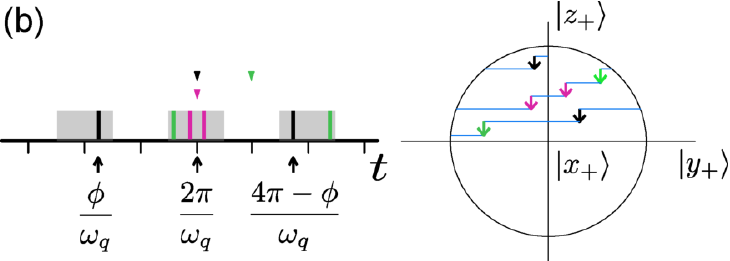}
    \caption{(Color online) SFQ pulse sequences. The shaded regions denote
        time
        windows
        in which SFQ pulses can induce a positive $y$-rotation.
        The right-hand panels depict the qubit trajectory on the
        front hemisphere ($\expval{x} > 0$) of the Bloch sphere. (a)
        Resonant pulse
        sequence. Here, pulses are applied at regular intervals of
        $2\pi/\omega_{q}$.
        The SFQ pulses induce small $y$-rotations spaced by free
        precession for a
        full period. (b) Symmetric pulse pairs. Three symmetric pairs
        are shown,
        denoted by the different colors. For the black pair, one pulse
        is delivered at
        $\phi/\omega_{q}$ and the other at $(4 \pi - \phi)/\omega_{q}$.
        For each pulse
        pair, the pulses are symmetric around $m\pi/\omega_{q}$
        (arrows). It is
        possible to overlay multiple symmetric pairs to induce larger
        qubit rotations.
        \label{fig:sequences}
    }
\end{figure}

In the simplest scheme for SFQ-based coherent control,
one applies a regular train of SFQ pulses that is synchronized to the qubit
oscillation period $T_q = 2\pi/\omega_q$ (Fig. \ref{fig:sequences}a);
this is the situation considered in \cite{McDermott2014}.
In this method, because
$U_{\rm fr}(T_c) = \hat{1}$ in the qubit subspace,
only the desired $y$-rotations can be induced by $U_{\rm fr}(T_c)$ and $U_{\rm
            SFQ}$; however, leakage out of the qubit subspace can
be significant. To attain
higher fidelity, it is necessary to clock SFQ pulses to the qubit at a higher
rate and to employ more sophisticated sequences.
When the qubit is no longer resonant with the SFQ clock,
$U_{\rm fr}(T_c) \neq \hat{1}$, and the time evolution of an arbitrary
sequence of $U_{\rm fr}(T_c)$ and $U_{\rm SFQ}$
is generally not confined to $y$-rotations.
To ensure high overlap with the target $y$-rotation, we construct a composite
sequence built up from \textit{symmetric pairs} of SFQ pulses delivered to the
qubit at times $\phi / \omega_{q}$ and $(2m \pi - \phi)/\omega_{q}$ for some
integer $m$, as shown in Fig. \ref{fig:sequences}b. The pulses of the symmetric
pair occur symmetrically with respect to time $mT_{q}/2$. We represent the
symmetric pair by the tuple notation $(m, \phi)$. As an example, we can write
the resonant sequence in terms of symmetric pairs:
the first and last pulses form the pair $(N_q, 0)$; the second and penultimate
pulses form the pair $(N_q, 2\pi)$, etc.
In general, the sequence can be described as the set of symmetric pairs $(N_q,
    2 \pi k )$ for each $k$ between $0$ and $N_q/2$.

\par To see that application of symmetric pulse pairs has the net effect of a
$y$-rotation within the
qubit subspace, we inspect the time evolution operator $U_{(m, \phi)}$
associated with symmetric pair $(m,\phi)$:
\begin{multline}
    U_{(m, \phi)} = U_{\rm fr}\left(\phi/\omega_{q}\right)
    U_{\rm SFQ}
    U_{\rm fr}\left((2m \pi-2\phi)/\omega_{q}\right)\\
    \times U_{\rm SFQ}U_{\rm fr}\left(\phi/\omega_{q}\right)
    \label{eqn:sym_construct}.
\end{multline}
We substitute $U_{\rm SFQ}$ from Eq.~\eqref{eqn:u_sfq_matrix},	expand to the
first order in $\delta\theta$, and obtain
\begin{align}
    U_{(m, \phi)} =
    \begin{bmatrix}
        1                         &
        -\cos(\phi) \delta \theta &
        0                           \\
        \cos(\phi) \delta \theta  &
        1                         &
        -\lambda \mu \delta \theta  \\
        0                         &
        \lambda \mu \delta \theta &
        \ldots
    \end{bmatrix} + {\cal{O}} (\delta \theta^2),
\end{align}
where
\begin{align}
    \mu = \exp\left(\frac{im \pi (2\omega_{q} + \omega_{21})}{\omega_{q}}
    \right)\cos\left(\frac{\omega_{21}}{\omega_{q}}(m
        \pi-\phi)\right).
    \label{eqn:u_sym}
\end{align}
We observe that, to first order in $\delta \theta$, $U_{(m, \phi)}$ is indeed a
$y$-rotation in the qubit subspace. Moreover, the dependence of leakage
on the timing of the symmetric pair through $\phi$ provides a degree of freedom
that will enable us to tailor subsequences in order to minimize leakage errors,
as we discuss in Sec. \ref{sec:SCALLOPS:LEAKAGE}. We remark that although
it is tempting to set $\mu=0$
by appropriate selection of $\phi$ and thereby eliminate the 1-2 transition,
the 0-1 transition will become very weak as a side effect.
In fact, there is an analogous composite microwave pulse method that exploits a
restricted form of this idea corresponding to $m=1$ \cite{Steffen2003};
however, the gate performance is no better than that of naive Gaussian pulses.

\par As we shall see below, the construct of symmetric pairs becomes
particularly advantageous when it is extended to the case of multiple symmetric
pairs $(m_i,\phi_i)$
applied at times $\phi_i/\omega_q$ and $(2 m_i \pi -\phi_i)/\omega_q$
for $i \in \mathrm{N}$, as shown in Fig. \ref{fig:sequences}b.
We note that the pulse pairs do interfere with each other
because they generally do not commute; however,
the resulting error is acceptably small for practical choices of
$\delta\theta$.

\subsection{Control of Qubits at Multiple Frequencies
    \label{sec:SCALLOPS:MULTIPLE}}

\par In general, the qubit oscillation period will not be commensurate with the
SFQ clock, so that the optimal delivery times of the symmetric pairs will not
exactly coincide with SFQ clock edges. As a result, it is necessary to round a
symmetric pair to a particular pair of clock edges $n_i$ and $n_j$.
We first note that $n_i$ and $n_j$ preserve the symmetry precisely if
the times at which the pulses are applied are symmetric with respect to
$mT_q/2$ for some
integer $m$:
\begin{equation}
    \frac{1}{2}\left(\frac{n_i}{N_c} \cdot \frac{2\pi N_q}{\omega_q} +
    \frac{n_j}{N_c} \cdot \frac{2\pi N_q}{\omega_q}\right) = m T_q/ 2 ,
    \label{eqn:precise_symmetry}
\end{equation}
where $N_c$ is the number of clock cycles and $N_q$ is the number of qubit
cycles in the sequence.
This condition is equivalent to the expression
\begin{equation}
    A_{\rm sym} = \left|\left(\frac{n_i}{N_c} N_q\right) \bmod 1 +
    \left(\frac{n_j}{N_c} N_q\right) \bmod	1 - 1 \right| = 0
    \label{eqn:approx_symmetry},
\end{equation}
where $A_{\rm sym}$ is now a measure of the violation of symmetry due to
mismatch between the SFQ clock and the qubit oscillation period. We find
empirically that coherent pulse errors are acceptably small for pulse pairs
delivered at times such that  $A_{\rm sym} < 0.05$. In the following, pulse
pairs that are termed symmetric are understood to satisfy this condition.

\par The delivery of SFQ pulses to the qubit as symmetric pairs constrains the
time evolution to the desired $y$-rotation; however, it is not obvious how to
control multiple qubits resonating at different frequencies, as demanded by the
surface code. For a qubit frequency that is not a subharmonic of the SFQ clock
frequency, the concern is that mismatch between the qubit oscillation period
and the SFQ clock will lead to phase error, as the precession of the qubit
during the gate is not an integer number of qubit cycles.

\begin{figure}[t!]
    \includegraphics[width=0.98\columnwidth]{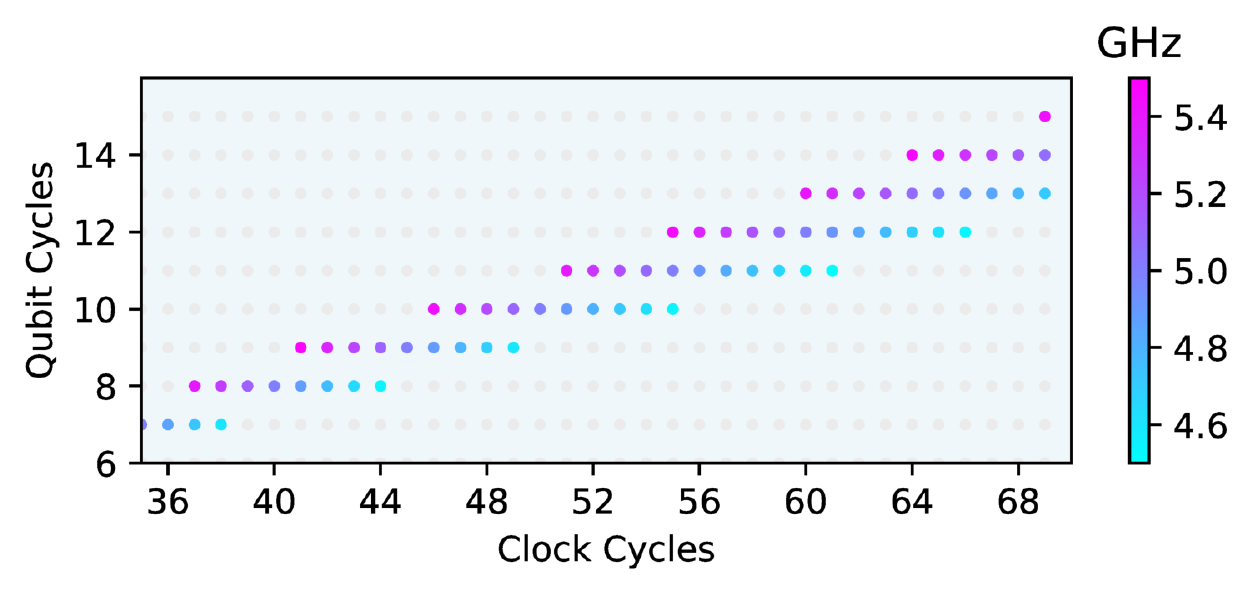}
    \caption{(Color online)
        Map of qubit frequencies permitting high-fidelity control.
        Each grid point represents a qubit oscillation frequency that
        satisfies
        Eq.~\eqref{eqn:matching}.
        The grid points highlighted with the color scale span qubit
        frequencies
        from $4.5$ to $5.5$ GHz, assuming an SFQ clock frequency of
        25~GHz.
        \label{fig:mismatch}}
\end{figure}

\par To avoid these phase errors, the key is to tune the qubit frequency
such that the total gate time $T_g$  corresponds to both an integer
number $N_c$ of clock cycles $T_c$ and an integer number $N_q$ of qubit
cycles $T_q$, so that $T_g = N_c T_c = N_q T_q$.
This relation translates into the following frequency matching condition:
\begin{align}
    \frac{N_q}{\omega_q} = \frac{N_c}{\omega_c},
    \label{eqn:matching}
\end{align}
where $\omega_c = 2\pi f_c$ is the angular frequency of the clock. We can use
Fig. \ref{fig:mismatch} to find frequencies that satisfy
Eq.~\eqref{eqn:matching}.
In this diagram, each grid point represents a qubit frequency
(shown in color scale) determined from the above matching
condition. Again, we consider a 25~GHz SFQ clock;
for a small range of frequencies around each of
the ``magic'' qubit operating points
given by Eq. \eqref{eqn:matching}, accurate qubit control is possible.
\begin{figure}[t!]
    \centering
    \includegraphics[trim={3.5cm 14cm 13.5cm 3cm}, clip]{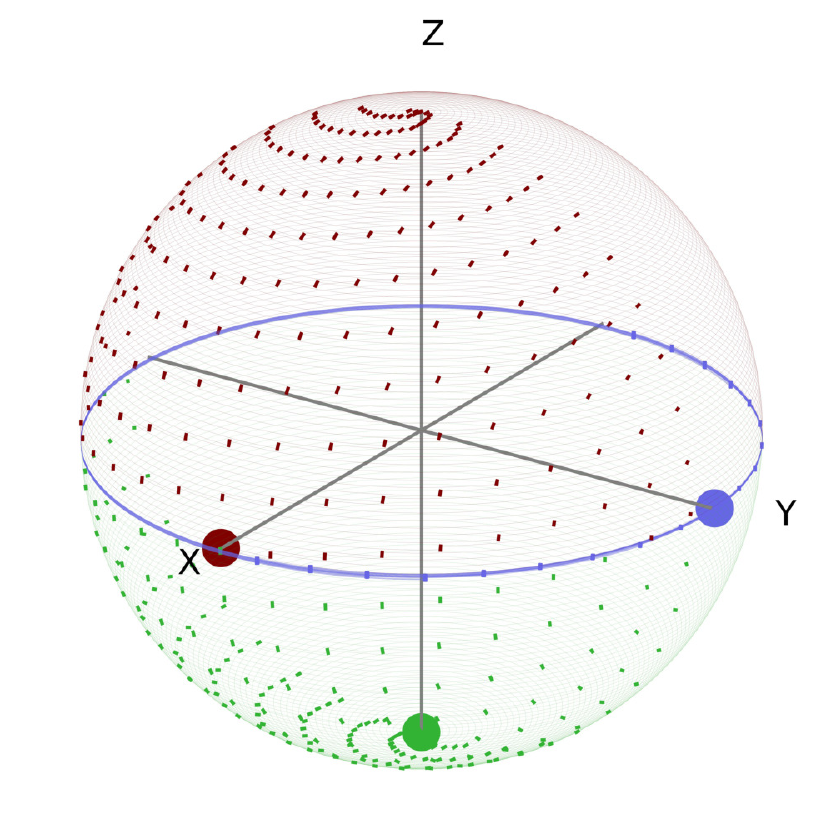}
    \caption{(Color online) Basic SCALLOP sequence. We show the trajectory
        on the
        Bloch sphere for qubits initialized in states $\ket{x_+}$
        (green), $\ket{y_+}$
        (purple), and $\ket{z_+}$ (red). Under the influence of the
        pulse sequence, the
        state $\ket{x_+}$ is rotated to $\ket{z_-}$; the state
        $\ket{y_+}$ undergoes
        small oscillations on the equator but returns to itself; and
        the state
        $\ket{z_+}$ moves to $\ket{x_+}$, as expected for a $Y_{\pi/2}$
        gate.
        \label{fig:base_scallop}}
\end{figure}

\par From frequency-matching relation \eqref{eqn:matching}, it is clear that
longer gate times will permit high-fidelity control of a larger number of
distinct qubit frequencies. However, the number of register bits needed to
describe the pulse sequence can be drastically reduced by the repeated
streaming of high-fidelity \textit{subsequences}. This strategy leads to
compact registers that are efficient to implement in hardware,
and it provides a desirable periodic suppression of leakage as a side effect,
as we discuss below in Sec. \ref{sec:SCALLOPS: VERIFICATION}.

\par In more detail, the length of the subsequence is a trade-off between the
number of register bits and the performance of the subsequence.
If the length is too short, the search space will be
too restricted to contain high-fidelity subsequences, and
the number of qubit frequencies we can
control will be decreased (see Fig. \ref{fig:mismatch}).
In our simulations, we find a good balance for subsequences
consisting of $35-55$ bits. The number of subsequence repetitions,
and thus the overall length of the gate, is set by the
size of the coherent rotation $\delta \theta$ imparted to
the qubit per SFQ pulse.
The relationship between gate time and tip angle is
$T_g \propto T_c/(2 \delta \theta)$, and it is tempting to reduce
the gate time by increasing $\delta \theta$; for large tip angle, however,
errors that are second order in $\delta \theta$ will become significant.
We find that $\delta \theta \approx 0.03$ is optimal,
corresponding to a reasonable coupling capacitance from the
SFQ driver circuit to the qubit island of order 100~aF for typical transmon
parameters. For the simulations described here, we target $Y_{\pi/2}$ gate time
around $12$~ns.

\par With this frequency matching condition and approach to hardware
optimization, we can immediately construct some \textit{basic subsequences} as
follows. Given the number of clock cycles $N'_c$ and qubit cycles $N'_q$ in a
subsequence, for each clock cycle $i \in [0, N'_c]$ we apply an SFQ pulse on a
given clock edge provided the pulse induces a rotation in the positive
$y$-direction. We then repeat the subsequence an appropriate number of times to
achieve the target rotation.
Explicitly, we deliver an SFQ pulse to the qubit on the $k^{th}$ clock
edge of the subsequence provided the following condition is
fulfilled:
\begin{equation}
    \left(N'_{q} \cdot \frac{k}{N'_{c}}\right) \bmod\, 1 \leq 1/4 \text{ or
    }
    \geq 3/4.
    \label{eqn:simple_pulse_criteria}
\end{equation}
This class of subsequences is expected to yield reasonably high fidelity
because it has a palindrome structure, which implies that pulses are delivered
to the qubit as symmetric pairs. For example, the first and last pulses form
the pair $(N_q, 0)$;
the second and penultimate pulses form the pair $(N_q, \omega_q T_c)$, etc.
In general, the sequence contains a pair
$(N'_q, k \omega_{q}T_c)$ for each $k$ between $0$ and $N'_c/2$ that satisfies
\eqref{eqn:simple_pulse_criteria}.

\par As an example, we simulate a sequence built from 10 repetitions of a basic
subsequence using $N'_c = 39$ and $N'_q = 8$; a plot of the qubit trajectory on
the Bloch sphere is shown in Fig. \ref{fig:base_scallop} for a $5.12781$~GHz
qubit initialized along the $+x$ (green), $+y$ (purple), and $+z$ (red)
directions. Here, the tip angle $\delta \theta= 0.0126$ is chosen to achieve
the $Y_{\pi/2}$ rotation in 390 clock steps.
Assuming a qubit anharmonicity of 250~MHz and a 25~GHz SFQ clock frequency,
this sequence achieves fidelity of $99.9\%$ in under $16$~ns.
Although this scheme for constructing basic subsequences demonstrates the
possibility of controlling multiple qubit frequencies using a single global
clock, it is by no means optimal, as the achieved fidelity is rather modest.
The dominant source of infidelity is leakage from the
computational subspace. In the following subsection, we describe an approach to
suppress this leakage.

\subsection{Leakage Suppression \label{sec:SCALLOPS:LEAKAGE}}

At the core of SCALLOPS is the optimization algorithm that
eliminates leakage from the computational subspace.
Starting with a \textit{basic subsequence}
of the type described in Sec. \ref{sec:SCALLOPS:MULTIPLE},
we need to flip bits in order to suppress
leakage while preserving the target rotation
in the qubit subspace. The major difficulty
in subsequence optimization is that bit
flips that reduce leakage will generally disrupt
the rotation in the qubit subspace. This problem
is analogous to solving a Rubik's cube: when
the cube mismatched at the top layer, a naive
set of operations to complete the top layer
will generally disrupt the other layers that
are already matched. This difficulty can be
circumvented by using a sequence of operations
whose net effect is felt only at the top layer.
We can solve the qubit control problem analogously:
the corresponding sequence of operations is to
flip a symmetric pair of bits in the subsequence
and to scale the tip angle $\delta \theta$ to preserve
rotation in the qubit subspace. While this latter
step might seem dubious, given that $\delta \theta$
is fixed by the geometric coupling of the SFQ driver
to the qubit, we will show that
for a given qubit frequency satisfying the matching
condition Eq. \eqref{eqn:matching} there exists a high
density of high-fidelity subsequences in the space
of tip angles $\delta \theta$. Our strategy will be
to allow $\delta \theta$ to vary as we search for a cluster of
high-fidelity, low-leakage subsequences.
Then we will select those subsequences that achieve highest fidelity for the
specific value of $\delta \theta$ dictated by the available hardware.
\begin{figure}[t!]
    \includegraphics{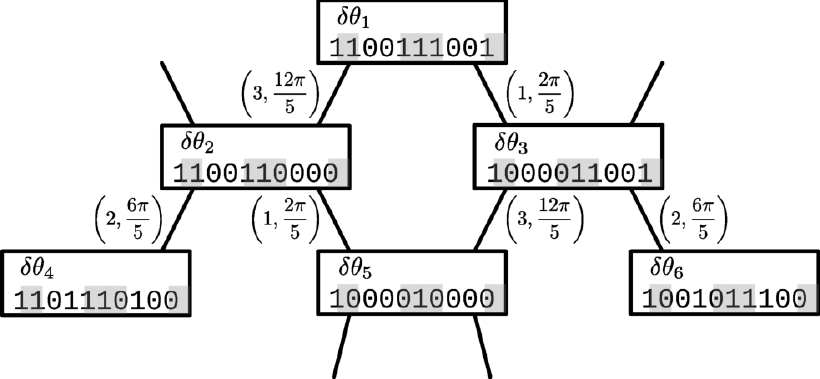}
    \caption{(Color online) Subsequence Graph. Each vertex represents a
        10-clock
        cycle (2-qubit cycle)
        subsequence with a distinct tip angle $\delta \theta_i$ in the
        qubit subspace
        (we consider a $25$~GHz clock and a $5$~GHz qubit). The shaded
        regions denote
        windows in which positive $y$-rotations can be induced by the
        application of
        SFQ pulses. Vertices that are connected differ by a single
        symmetric pair
        $(m,\phi$), which labels the connection. For example, the
        subsequence at the
        top of the graph differs from its left neighbor by the pair
        $(3, 12\pi/5)$,
        corresponding to pulses applied on the sixth and ninth clock
        edges following initiation of the sequence (clock edge zero).
        \label{fig:graph}}
\end{figure}

\begin{figure*}
    \centering
    \includegraphics[width=0.98\textwidth]{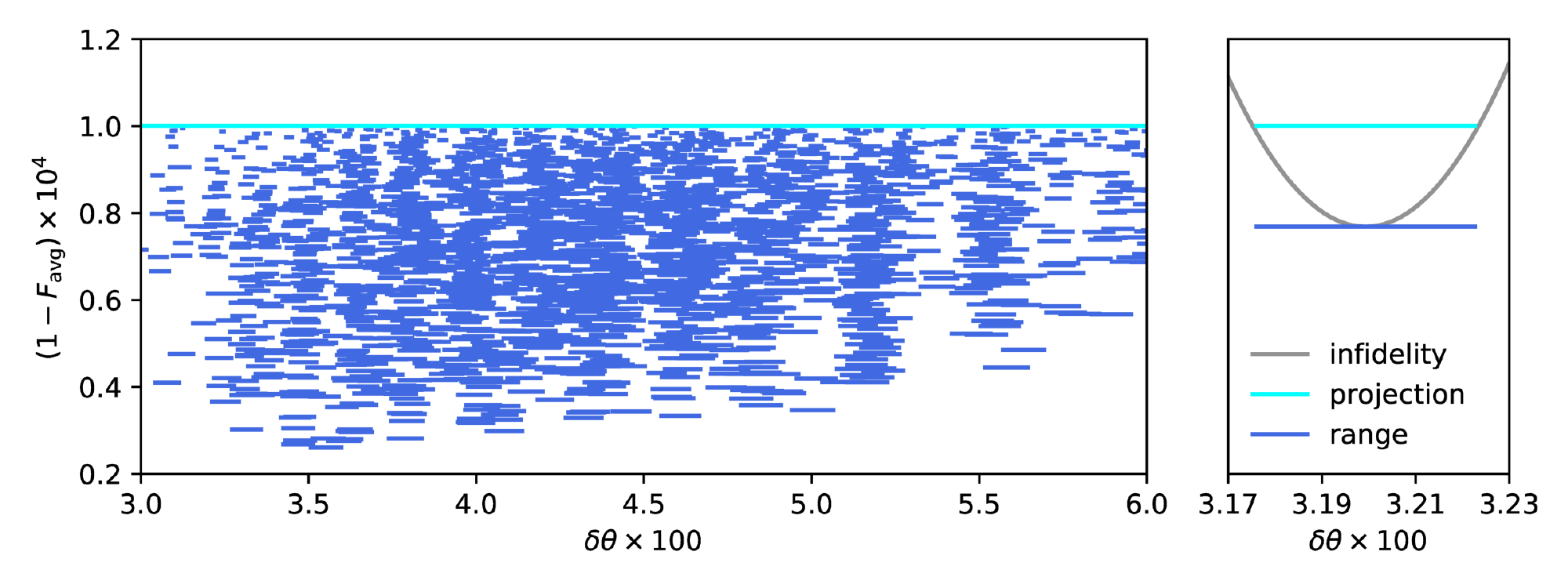}
    \caption{(Color online) Gate infidelity versus tip angle $\delta
        \theta$.
    Each horizontal bar represents a unique subsequence. The bars are
    centered
    horizontally at the
    optimal tip angle $\delta \theta_{\rm opt}$, and the vertical position
    of the
    bars represents the minimum subsequence infidelity. The horizontal
    extent of
    each bar denotes the range of $\delta \theta$ over which the infidelity
    of the
    subsequence remains below $10^{-4}$ (see inset).
    The cyan trace is the projection of all subsequences onto the line
    $1-F_{\rm
        avg}=10^{-4}$; these segments merge into a nearly continuous
    line that spans
    the range of tip angles from 0.03 to 0.06 rad.
    \label{fig:neighborhood}
    }
\end{figure*}
\begin{figure*}[t!]
    \centering
    \includegraphics[width=0.98\textwidth]{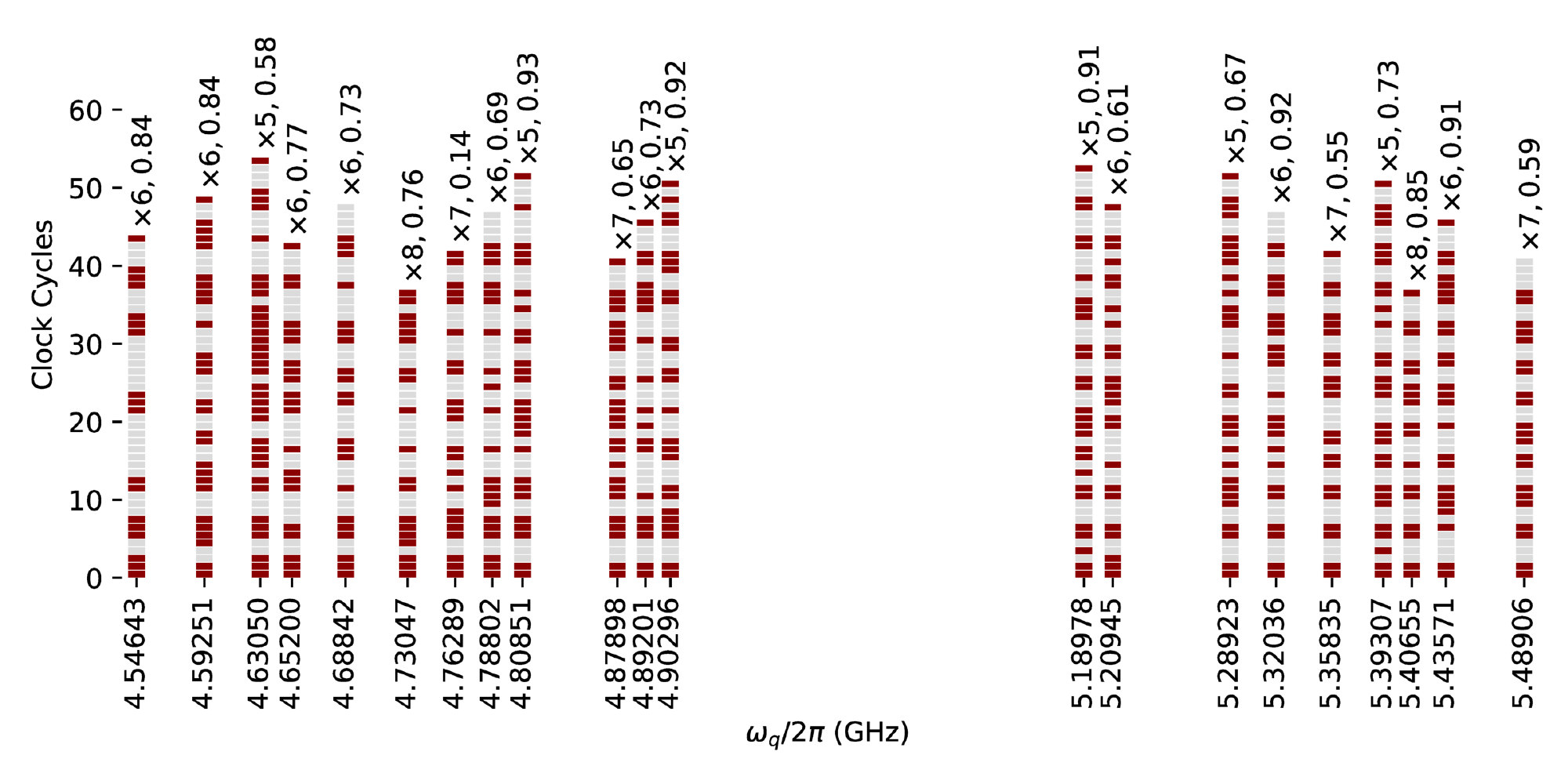}
    \caption{(Color online) A collection of SCALLOP subsequences. The
        subsequences
        are labeled below with the frequency of the target qubit and
        above with the
        achieved gate fidelity (in units of $10^{-4}$) and number of
        repetitions
        required to achieve the $Y_{\pi/2}$ gate.
        Here, time flows upward; red (grey) bars correspond to clock
        cycles during
        which SFQ pulses are applied (omitted). The SCALLOP sequences
        span $21$ qubit
        frequencies and share a fixed tip angle $\delta\theta = 0.032$.
        The frequency spacing of the subsequences is slightly adjusted
        for improved
        readability.
        \label{fig:subsequences}}
\end{figure*}
\par More formally, we can describe this method in terms
of a subsequence graph $G = (V, E)$, where the vertices $V$ represent
individual SFQ subsequences with their optimal tip angles $\delta\theta$
and the connections $E$ link subsequences that are separated
by a single symmetric pair of bit flips. Explicitly,
\begin{itemize}
    \item Each vertex $V$ is described by a subsequence bit pattern
          $\mathbf{S}$
          and its optimal tip angle $\delta \theta_{\rm opt}=
              \text{argmax}_{\delta\hat{\theta}}F_{\rm avg}$.
    \item Each connection $E$ links subsequences  $(\mathbf{S},
              \mathbf{L})$ that
          differ by a single symmetric pair $(m,\phi)$.
\end{itemize}
We define $V$ and $E$ in this way with the
goal of separating control in the qubit subspace from
leakage elimination: navigation through the subsequence graph $G$
preserves rotation in the qubit subspace, but movement from vertex to vertex
can change leakage out of the computational subspace substantially, as one can
see from Eq. \eqref{eqn:u_sym}. A trivial example of the subsequence graph is
shown in Fig. \ref{fig:graph}.

\par With this definition of the subsequence graph $G$, we describe
a simple procedure to find high-fidelity subsequences. We first construct
a \textit{basic subsequence} $\mathbf{S}$ as defined in
Sec. \ref{sec:SCALLOPS:MULTIPLE}. This subsequence serves
as the entrance point to the subsequence graph. We then explore all vertices
adjacent to $\mathbf{S}$ and greedily move to the vertex with the highest
fidelity. We repeat this greedy move until we reach a local
fidelity maximum. This typically takes only 5-10 steps.
In general, 5-8 repetitions of such high-fidelity subsequences will yield gates
with fidelity greater than $99.99\%$ in a total sequence time under $12$~ns.

\par The subsequences generated by the algorithm described above
are not yet sufficient for experimental implementation because we have allowed
the tip angle per SFQ pulse to vary during our search. In practice, the tip
angle is determined by the coupling capacitance of the SFQ driver to the
transmon qubit and cannot be exquisitely controlled during fabrication, or
varied \textit{in situ} following fabrication.

\par The solution to the problem is to explore a larger region of the
subsequence graph and  to identify a large ensemble of high-fidelity candidate
subsequences
corresponding to a range of optimal tip angle $\delta \theta_{\rm opt}$. For
each of these subsequences, high-fidelity rotations (say, infidelity under
$10^{-4}$) are achieved over a range of $\delta \theta$, so that it is
straightforward to identify from this ensemble specific subsequences that yield
high fidelity for a fixed $\delta \theta$. More specifically, we ignore all
vertices with fidelity
lower than $99.99\%$ and perform a standard breadth-first search to
traverse the remaining vertices of the graph, leading to a set of characterized
subsequences which we call the subsequence \textit{neighborhood}.
\begin{figure}[t!]
    \centering
    \includegraphics[width=0.45\textwidth]{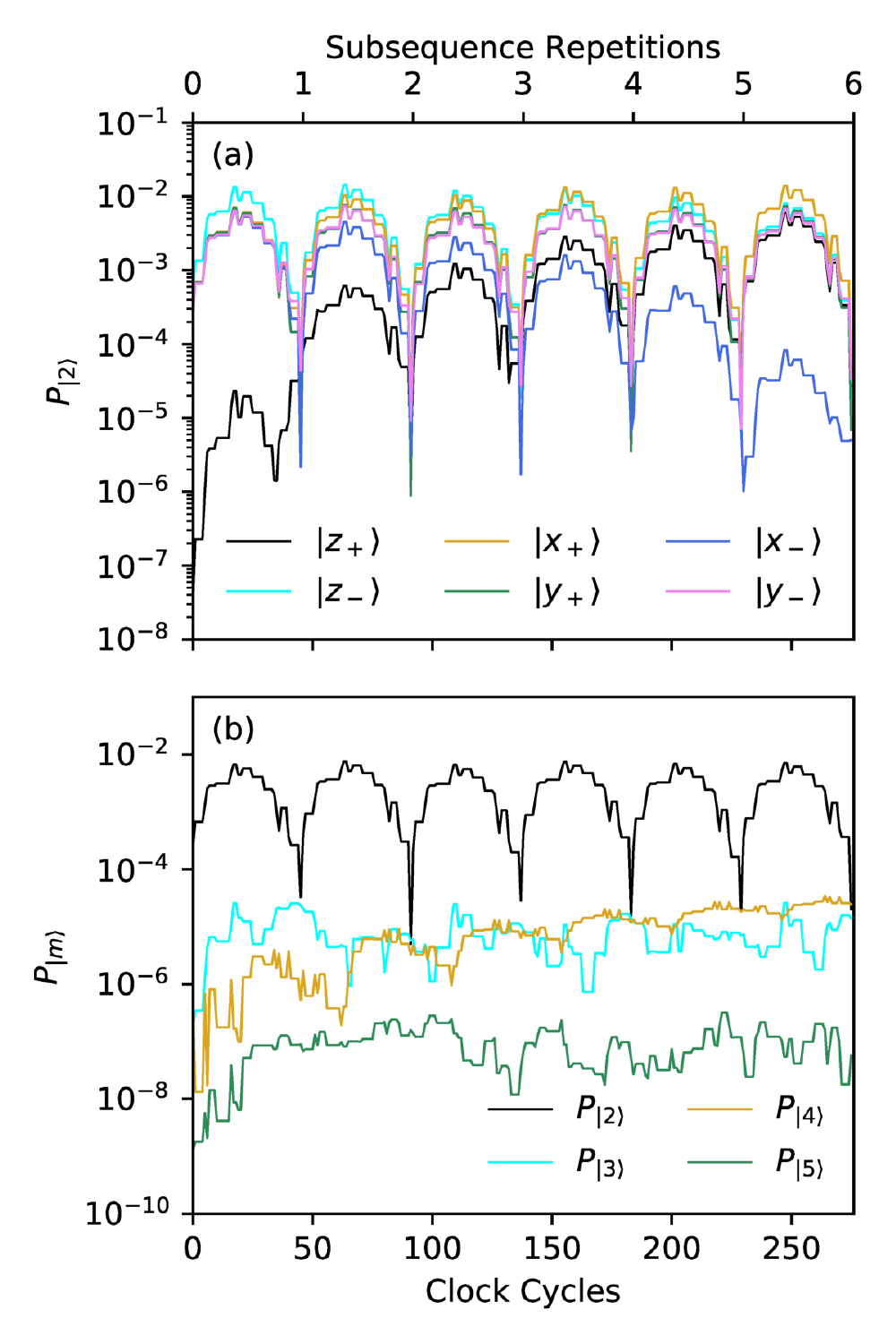}
    \caption{(Color online) Leakage into  noncomputational states for the
        sequence of Fig. \ref{fig:neighborhood}
        corresponding to the 4.89201~GHz qubit. The sequence involves 6
        repetitions
        of a subsequence consisting of 46 bits. (a) Population of
        $\ket{2}$ for initial
        qubit states aligned along the cardinal directions of the Bloch
        sphere. (b)
        Population of states $\ket{2}$, $\ket{3}$, $\ket{4}$, and
        $\ket{5}$ averaged
        over the same initial qubit states.}
    \label{fig:leakage}
\end{figure}
\begin{figure}[t!]
    \centering
    \includegraphics[width=0.45\textwidth]{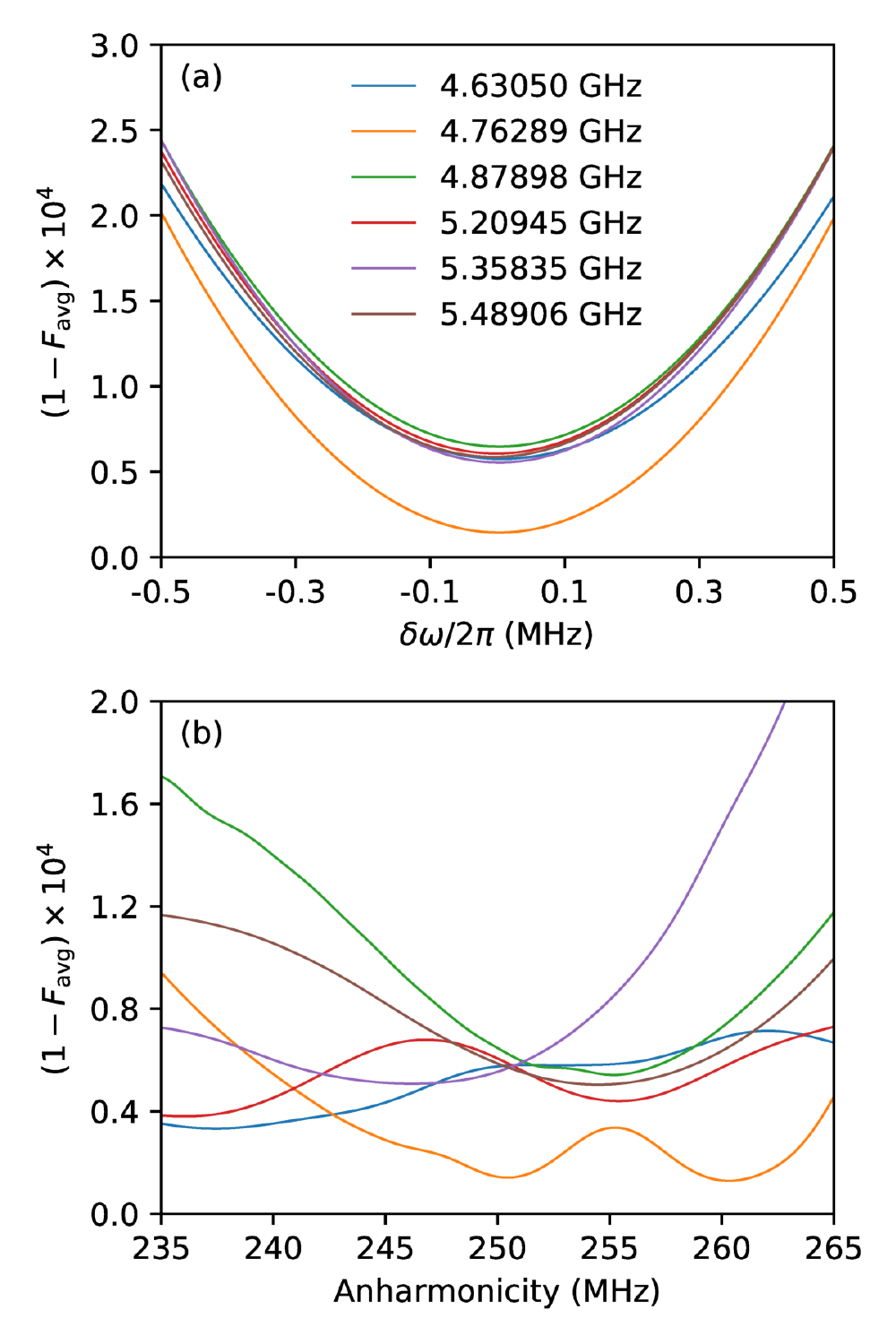}
    \caption{(Color online) Sensitivity of gate fidelity to variation in
        (a)
        qubit frequency and (b) anharmonicity. The simulation is run
        for six different
        qubit frequencies.}
    \label{fig:sensitivity}
\end{figure}

\par In Fig. \ref{fig:neighborhood} we plot the infidelities achieved versus
the tip angle $\delta \theta$ for a subsequence neighborhood associated with a
$4.65200$~GHz qubit. For each individual subsequence in the neighborhood, high
fidelity is reached for only a small range of tip angles around the optimal
value. However, given a fixed value of $\delta \theta$, numerous subsequences
are available that achieve gate fidelity well beyond the target of 99.99\%.
This is true for SFQ tip angle spanning a broad range from $0.03$ to $0.06$,
which is more than enough to accommodate any inaccuracy in the design of the
coupling capacitance between
the SFQ driver and the transmon.

\subsection{Sequence Verification \label{sec:SCALLOPS: VERIFICATION}}

\par We have performed the above-described neighborhood search for $21$
different
frequencies satisfying the matching condition given by
Eq.~\eqref{eqn:matching}.
The result is shown in Fig. \ref{fig:subsequences}.
While a 3-level model of the transmon was used to derive the sequences, the
presented fidelities were calculated for a model incorporating 7 energy levels.

\par In Fig. \ref{fig:leakage}, we examine leakage into the noncomputational
states $\ket{2}$, $\ket{3}$, $\ket{4}$, and $\ket{5}$ for the SCALLOP sequence
corresponding to the 4.89201~GHz qubit. We observe that the dominant leakage
into state $\ket{2}$ is roughly bounded at $10^{-2}$ for initial qubit states
spanning the cardinal points on the Bloch sphere. Moreover, as the qubit state
approaches $\ket{0}$, we see that leakage into $\ket{2}$ is particularly low,
as demonstrated in the curves corresponding to initial
states $\ket{z_+}$ and $\ket{x_-}$. While the population of state $\ket{2}$ can
approach $10^{-2}$ toward the middle of the subsequence, the population always
drops below $10^{-4}$ at the completion of each subsequence repetition, as the
subsequences are explicitly constructed to minimize leakage from the qubit
subspace. The  population of states
$\ket{3}$ and $\ket{4}$ is well below $10^{-4}$ throughout, while states
$\ket{5}$ and higher have negligible populations.

\par Finally, in Fig. \ref{fig:sensitivity} we simulate the effect of qubit
parameter variation on SCALLOP gate fidelity.
Error from frequency drift can be modeled as an ideal gate followed by a small
precession:  $U_{\rm fr}(\delta \omega T_g/ \omega_{q})Y_{\pi/2}$.
From Eq. \eqref{eqn:fid}, the infidelity of this gate is then approximately
$(\delta \omega T_g)^2/6$. For gate fidelity to degrade by $10^{-4}$,
the qubit frequency drift $\delta \omega/2\pi$ must reach about 300~kHz, given
a gate time of 12~ns. This naive estimate is in qualitative agreement with the
full simulation results in Fig. \ref{fig:sensitivity}a; note that based on the
above argument, we expect microwave-based qubit gates to display similar
sensitivity to qubit frequency drift. In Fig. \ref{fig:sensitivity}b we see
that SCALLOP gate fidelity is relatively insensitive to variation in qubit
anharmonicity. In a practical system, the anharmonicity of each qubit would be
calibrated upon system bringup. As anharmonicity is set by the transmon
charging energy, it is not expected to fluctuate in time.

\section{CONCLUSION\label{sec:CONCLUSION}}

\par We have performed numerical simulations to demonstrate coherent qubit
control across multiple frequencies using irradiation with classical bits
derived from the SFQ logic family. Using a single global clock at 25~GHz to
stream pulses from compact registers consisting of 35-55 bits, we achieve gate
fidelity better than $99.99\%$ across $21$ qubit frequencies spanning the range
from 4.5 to 5.5~GHz. The control subsequences are readily amenable to storage
in compact SFQ-based shift registers, as outlined in \cite{McDermott2018}. We
have described an intuitive, efficient method for the derivation of
high-fidelity SFQ-based pulse sequences that is readily adapted to arbitrary
single-qubit gates. The SCALLOPS method is robust in the sense that large
imprecision in the tip angle per SFQ pulse is readily accommodated by
appropriate variation in the subsequence bitstream.  The control approach is
immune to wiring parasitics and offers the possibility for tight integration of
a large-scale quantum array with a proximal classical coprocessor for the
purposes of reducing system footprint, wiring heatload, and control latency.

\vspace{12pt}

\begin{acknowledgments}

    This work was supported by the NSF under Grant QIS-1720304.

\end{acknowledgments}

\bibliography{scallop}
\bibliographystyle{apsrev4-1}
\end{document}